\documentclass[letterpaper, 10 pt, conference]{ieeeconf}  

\IEEEoverridecommandlockouts                              
\pdfoutput=1
\usepackage[utf8]{inputenc}
\usepackage[T1]{fontenc}
\usepackage[ruled, vlined]{algorithm2e}

\usepackage{graphicx}
\usepackage{amsmath}
\usepackage{multirow}
\usepackage[table,xcdraw]{xcolor}
\usepackage{caption}
\usepackage{subcaption}
\usepackage{hyperref}
\usepackage{booktabs}

\title{\LARGE \bf
OpenCSI: An Open-Source Dataset for Indoor Localization Using CSI-Based Fingerprinting
}

\author{Arthur Gassner, Claudiu Musat, Alexandru Rusu and Andreas Burg}

\begin{document}

\maketitle
\thispagestyle{empty}
\pagestyle{empty}

\begin{abstract}

Many applications require accurate indoor localization. Fingerprint-based localization methods propose a solution to this problem, but rely on a radio map that is effort-intensive to acquire.

We automate the radio map acquisition phase using a software-defined radio (SDR) and a wheeled robot. Furthermore, we open-source a radio map acquired with our automated tool for a 3GPP Long-Term Evolution (LTE) wireless link. To the best of our knowledge, this is the first publicly available radio map containing channel state information (CSI). Finally, we describe first localization experiments on this radio map using a convolutional neural network to regress for location coordinates.\\


\end{abstract}

\section{INTRODUCTION}
\subsection*{PREAMBLE}

The range of applications requiring accurate localization has been growing rapidly (warehouse management, autonomous driving, asset tracking). While solutions based on a \textit{Global Navigation Satellite System} (GNSS) -- such as the \textit{Global Positioning System} (GPS) -- offer a suitable solution for many outdoor use cases, their accuracy degrades greatly once the line-of-sight assumption is broken, which is the case in indoor environments and urban canyons.

In such environments, the propagation of the radio waves is subject to multipath, shadowing and channel fading. Fingerprint-based localization methods based on detailed channel state information (CSI) attempt to exploit the richness brought by those effects to solve the localization problem. Such methods work by finding a mapping between a location-dependent feature (called a fingerprint) and the location of the device. This fingerprint is extracted from the communication between the device to locate and one (or several) \textit{Access Points} (APs), such as WiFi modems or Long-Term Evolution (LTE) towers.

Fingerprint-based methods consist of two consecutive phases: the radio map acquisition phase and the training phase. In the radio map acquisition phase, the area in which the device needs to be localized is surveyed. That is, several locations in space are selected (called \textit{Reference Points} (RPs)), and the fingerprints recorded at those RPs are stored -- together with the coordinates of the RPs -- in a database. In the training phase, that radio map is fed to some machine learning model. The model learns to map fingerprints to their respective locations.

\begin{figure}[thpb]
  \centering
  \includegraphics[width=.49\textwidth]{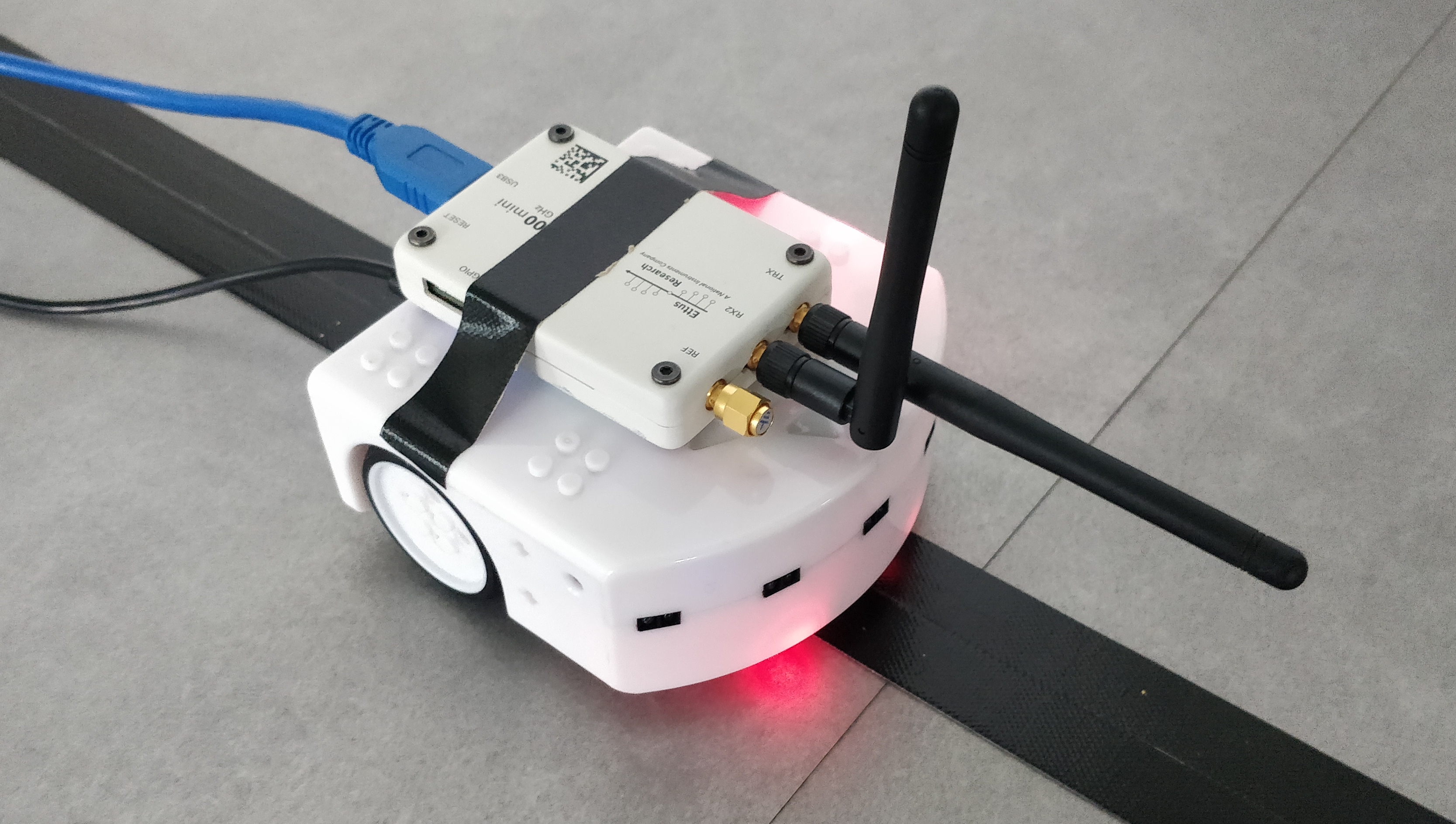}
  \caption{Software-defined radio (USRP B200mini) mounted on a wheeled robot (Thymio II \cite{thymio}), measuring the fingerprints along a line}
  \label{fig:thymio}
\end{figure}

In order to assess the performance of a fingerprint-based localization method, additional locations in space are selected (called \textit{Test Points} (TPs)), and the fingerprint at those locations are recorded. They are distinct from the RPs, and the fingerprints stemming from those TPs are used as a test set to evaluate how the method performs.

Fingerprint-based methods make the assumption that the fingerprint is location-dependent and that it changes through space in a manner that can be inferred by a model. Additionally, the fingerprint needs to be rich enough to uniquely identify the corresponding location.

A popular channel characteristic used as fingerprint is the \textit{received signal strength indicator} (RSSI) \cite{horus} \cite{cnn-rss-timeseries} \cite{wideep}. The RSSI represents an estimation of the power level that the receiver is getting from the AP. In empty space, the RSSI will decrease as the receiver moves away from the AP. However, in an indoor environment, subject to multipath and shadowing, this basic relationship does not hold anymore. Furthermore, the RSSI is a scalar and therefore several AP links are required in order to increase the dimensionality of the fingerprint. This measure is necessary to provide sufficient diversity to be able to localize the device accurately. 

Another channel characteristic -- the \textit{channel state information} (CSI) -- gained popularity as a fingerprint due to its high dimensionality \cite{fifs} \cite{deepfi} \cite{confi}. The CSI represents an estimate of the channel between the receiver and the AP. Its higher dimensionality (compared to the RSSI) allows for a finer representation of the radio waves' propagation environment between an AP and the device. This allows CSI-based fingerprinting methods to work with only one AP.

In the context of LTE, the CSI -- or \textit{channel estimate} is obtained periodically by the channel estimation in the receiver, and is typically expressed as the Fourier transform of the \textit{channel impulse response} (CIR). In the case of a channel bandwidth of 20MHz, the frequency domain representation of this impulse response is divided into 1200 tones (subcarriers), which are estimated based on a pre-defined training sequence every millisecond \cite{lte-in-a-nutshell}.

Hence, the CSI consists of 1200 complex numbers, and is defined as:

$$H_i = |H_i|e^{j\angle H_i} \qquad \text{for $i$ = 1,...,1200}$$

\noindent where $e$ is Euler's number, $i$ is the subcarrier index, $j$ is the complex unit, $|H_i|$ is the amplitude of the $i$-th subcarrier and $\angle h_i$ is the phase of the $i$-th subcarrier.

\subsection{LITERATURE REVIEW}

Most works studying fingerprinting are based on WiFi. One of the earliest attempts at fingerprinting -- called \textit{Horus} -- takes a Bayesian approach \cite{horus}. The RSSI of several WiFi APs are aggregated together and used as a fingerprint. In the test phase, the fingerprint is compared against the fingerprints in the radio map. For each RP in the radio map, the model predicts a probability for the fingerprint to match the location of the RP. The RPs and their predicted probabilities are then aggregated together into one final predicted location. To evaluate Horus' performance, two radio maps are acquired. One is built on the floor of a building (68.2m x 25.9m) with 21 WiFi APs covering it. In total, 172 RPs are gathered. The other radio map is built in an office space (11.8m x 35.9m) with 6 WiFi APs covering it. In total, 110 RPs are gathered. In both radio maps, 100 TPs are gathered on a different day, and both the RPs and TPs are spread evenly across the space. On both radio maps, Horus achieves a median error of 0.5m. The main downside of this method is the need for a large amount of APs, due to the fingerprint being made out of multiple RSSI measurements.

In \cite{fifs}, another Bayesian approach -- called \textit{FIFS} -- is taken, similar to the one in Horus \cite{horus}. This is the first work to take advantage of CSI. The CSI of several APs are aggregated together to form a fingerprint. To evaluate FIFS' performance, two radio maps are acquired. One radio map is built in a laboratory (7m x 11m), with 3 WiFi APs located within the laboratory. A total of 40 RPs are arranged in a grid (spacing of 1.2m). The other radio map is built in a corridor (32.5m x 10m), with 6 WiFi APs covering it. In total, 28 RPs following some kind of path along the corridor are gathered (spacing of 2m). FIFS achieves a mean error of 0.65m in the laboratory and 1.07m in the corridor. Moreover, FIFS is shown to outperform Horus on those same radio maps. However, the locations of the TPs -- which can greatly influence a method's performance -- are unspecified.

In \cite{deepfi}, a deep-learning approach -- called \textit{DeepFi} -- is taken in order to map the fingerprints to their location. A fingerprint is composed of the amplitude of the CSI frequency components extracted from one WiFi AP. The model consists of one stacked autoencoder (SAE) per RP, each trained to reconstruct the fingerprint attached to that RP. In the test phase, the fingerprint is fed to each SAE, and their reconstruction errors are aggregated to output a predicted location. To evaluate DeepFi's performance in different scenarios, two radio maps are acquired. One is gathered in a living room (4m x 7m), with 50 RPs arranged in a grid (spacing of 0.5m). Out of the 50 RPs, two lines (12 locations) are selected to be used as TPs, while the remaining locations are used as RPs. The other radio map is gathered in a laboratory (6m x 9m), with 50 RPs and 30 TPs, arranged evenly across the space not occupied by the desks in the room. In the living room and the laboratory, DeepFi achieves a median error of 0.7m and 1.4m, respectively. Moreover, DeepFi outperforms FIFS and Horus on both radio maps. This work shows that sub-meter median error can be achieved with only one AP when exploiting the fine-grain information held in the CSI. However, it has the drawback of requiring to train, store and do a forward pass through as many neural networks as there are RPs.

The complexity issue of DeepFi is addressed in \cite{confi}, where a convolutional neural network (CNN) approach -- called \textit{ConFi} -- is taken. The fingerprints are represented as images, and contain the amplitude of the CSI frequency components extracted from the communication with one WiFi AP over a certain measurement time window. The AP is transmitting on three different antennas, allowing the receiver to extract one CSI per antenna. Additionally, the model captures the time-dependency of the fingerprint by aggregating consecutive CSI measurements in an image. To evaluate ConFi, one radio map is acquired in an indoor space (16.3m x 17.3m) with several rooms. The 64 RPs are evenly spread (spacing between 1.5m and 2m). Those RPs are then split in RPs, TPs and locations used for validation. ConFi achieves a mean error of 1.36m on this radio map, which slightly outperforms DeepFi (mean error of 1.49m). However, ConFi does not suffer from the complexity issue of DeepFi as it consists of only one single NN.

All the above works are performing fingerprint-based indoor localization using WiFi. However, some effort to implement similar solutions using the LTE network were also successful \cite{lte-0} \cite{lte-1} \cite{lte-2}. A notable difference to WiFi is that the implementation of LTE allows for a better frequency resolution due to substantially more subcarriers, which in turn provides higher dimensional input features for the localization. However, we note that the amount of information that is present in a CSI measurement is ultimately limited by the signal bandwidth which for LTE is comparable to that of WiFI (20 MHz).

In \cite{lte-0}, a fingerprint-based method using one LTE eNodeB (located outdoor) is devised. 11 channel parameters calculated from the \textit{channel impulse response} (CIR) -- which corresponds to the result of passing the CSI through an inverse fast Fourier transform (IFFT). A subset of those channel parameters are selected by a feature-extraction algorithm and are then used as a fingerprint. A 3-layers fully-connected neural network regresses the fingerprint to the (x,y) coordinates of their RP. To test their model, an indoor radio map is built on the ground-floor (21m x 28m) of an empty building. In total, 58 RPs are gathered (spacing between 1m and 3m), following some sort of path. On this radio map, the model achieves a median error of 6.65m. The locations of the TPs -- which can greatly influence a method's performance -- is however not clear from the publication.

In \cite{lte-1}, a CSI-based fingerprinting method is described, where the device to locate communicates with several LTE eNodeBs located outdoor. Said fingerprints are made out of different descriptors of the CSI's shape, as well as RSSI and \textit{reference signal received power} (RSRP). The radio map is acquired indoor, in an apartment. A total of 113 RPs (spacing of roughly 0.5m) are gathered following some kind of path through the apartment. The TPs are seemingly randomly sprinkled throughout that path. The best model presented in the paper achieves a median error of 1.94m. Unfortunately, no comparison with previous state-of-the-art performances is made and the radio map remains closed-source, making result replication difficult.

In \cite{lte-2}, the amplitude of the CSI of one LTE eNodeB is used as fingerprint. The AP broadcasts on several antennas, allowing the extraction of several antenna-specific CSI measurements per measurement. Similarly to \cite{confi}, their model attempts to extract information from the amplitude fluctuation of the fingerprint through time. One indoor radio map is acquired (3.6m x 6m), with 25 RPs placed in an evenly-spaced grid (spacing of 1.2m), and 15 TPs placed seemingly randomly near the RPs. On this radio map, the model achieves a median error of 0.5m. Unfortunately, also for this algorithm the radio map is once more not made publicly available.

\subsection{CONTRIBUTIONS}

The first step of devising a fingerprint-based localization method is the radio map acquisition. When comparing the performance of different localization methods, it is necessary to use the same radio map. Indeed, the environment in which the radio map is acquired will impact the ease with which a model will map a fingerprint to its corresponding RP. Additionally, characteristics of the radio map such as the distance between the RPs, the amount of RPs used for training or the positions of the TPs relative to the RPs will affect the performance of the almost any algorithm. As a result, the lack of publicly available radio map forces researchers to acquire the radio map from scratch, as well as re-implement state-of-the-art localization methods to test them on their specific radio maps. To alleviate this issue, this paper provides the following contributions:

\begin{itemize}
    \item We provide an automated way to aquire a radio map, using readily available hardware. The method consists of a software-defined radio (SDR) mounted on a wheeled robot. The robot follows a line on the ground and stops periodically to record the fingerprint. The source-code controlling the robot\footnote{Available here: \url{https://github.com/arthurgassner/thymio-radio-map}} and the software-defined radio\footnote{Available here: \url{https://github.com/arthurgassner/srsLTE}} are open-sourced.
    
    \item We open-source a radio map for CSI-based fingerprinting. This radio map can serve as a basis for developing algorithms without the need to acquire first a radio map. It also facilitates comparison of different algorithms that are trained and tested on the same radio map (in the same environment).
    
    \item We perform a first analysis on our open-sourced radio map by implementing a rudimentary CNN-based model and evaluate it on our open-source radio map.
\end{itemize}

\subsection{OUTLINE}

In Section~\ref{section:radio-map-construction-methodology}, the methodology to automate the radio map acquisition phase is explained. In Section~\ref{section:open-sourced-radio-map}, the details regarding the open-sourced radio map are laid out. In Section~\ref{section:analysis}, an analysis of the radio map is performed using a rudimentary CNN-based localization method. In Section~\ref{section:experimental-results}, the results of our model on the open-sourced radio map are presented.

\section{RADIO MAP ACQUISITION METHODOLOGY} \label{section:radio-map-construction-methodology}

The radio map acquisition is a tedious and repetitive task. Furthermore, to acquire a dense radio map (with small spacing between RPs) often represents a prohibitive effort that cannot be performed manually.

\begin{figure}[thpb]
  \centering
  \includegraphics[width=.49\textwidth]{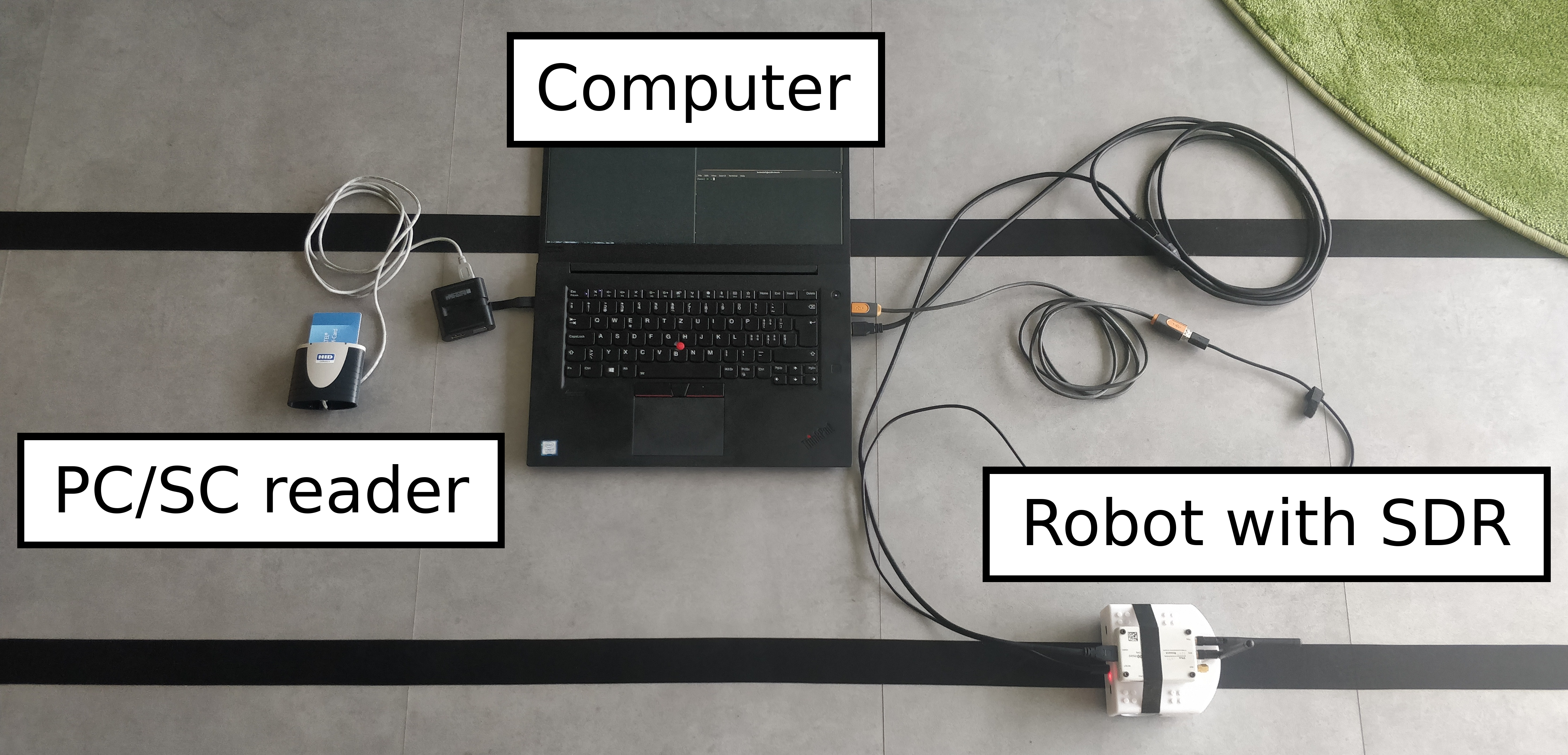}
  \caption{Overall setup}
  \label{fig:overall-setup}
\end{figure}

To automate this task, a robot mounted with a software-defined radio (SDR) (see Fig. \ref{fig:overall-setup}) is used in conjunction with a line-following algorithm to gather the fingerprints. The trajectory that must be followed by the robot is outlined on the ground by black tape. The robot then follows the line, stopping periodically to record a certain amount of consecutive fingerprints.

The overall setup (see Fig. \ref{fig:overall-setup}) comprises a computer running Ubuntu 18.04, a wheeled robot, an SDR and a \textit{personal computer/smart card} (PC/SC) reader. Everything is connected by USB to the computer. The PC/SC reader accesses a SIM card to authenticate the SDR when connecting to the LTE eNodeB.

The wheeled robot -- a \textit{Thymio II} \cite{thymio} -- was chosen for its availability, affordability and ease of use. IR proximity sensors placed under the robot allow for the implementation of a line-following behavior. Dead-reckoning keeps track of the position of the robot along the line. The robot is connected to a computer running the main control loop. The odometry error of the robot is estimated by having it follow a straight line (5m long). Over 10 runs, the maximum absolute difference between the believed travelled distance and observed distance is 25cm (5\% of the target distance).

The SDR -- a \textit{USRP B200mini} -- is mounted on the Thymio~II and is used to gather the fingerprints at the different locations (see Fig. \ref{fig:thymio}). The SDR is connected to a computer running a modified version of a free and open-source LTE software suite, called \texttt{srsLTE} \cite{srslte}.

Our slightly modified version of \texttt{srsLTE} allows to extract the CSI, alongside other channel-specific informations. The modified software writes data to three files. All three files are updated every time a channel estimate is performed by the \texttt{srsLTE} software, i.e. every millisecond. The first file contains the CSI of each antenna port used by the eNodeB, each in the form of 1200 complex numbers, corresponding to the frequency domain channel estimate. The second file contains the RSRP [dBm], RSSI [dB], \textit{reference signal received quality} (RSRQ) [dB], \textit{signal-to-noise Ratio} (SNR) [dB], \textit{carrier frequency offset} (CFO) [kHz] and noise estimate [dBm]. Finally, the third file contains details about the connection: \textit{physical cell identifier} (PCI), the amount of \textit{physical resource block} (PRB) sent over the channel (which can be inferred from the bandwidth of the channel) as well as the amount of ports and Rx antennas. In all three files, the index of the \textit{transmission time interval} (TTI) -- corresponding to 1ms -- is written alongside the newly written data as a way to timestamp the appended information.

The trajectory that must be followed by the robot is outlined on the ground by black tape. The robot then follows the line, stopping every time a user-specified distance has been traveled to record a user-specified amount of consecutive CSI fingerprints.

\section{OPEN-SOURCED RADIO MAP} \label{section:open-sourced-radio-map}

In order to facilitate the development and comparisons of CSI-based localization methods, a radio map is acquired and made publicly available \footnote{The dataset can be downloaded through the following link: \url{https://figshare.com/articles/dataset/openCSI/19596379}}.

\begin{figure}[thpb]
  \centering
  \includegraphics[width=.49\textwidth]{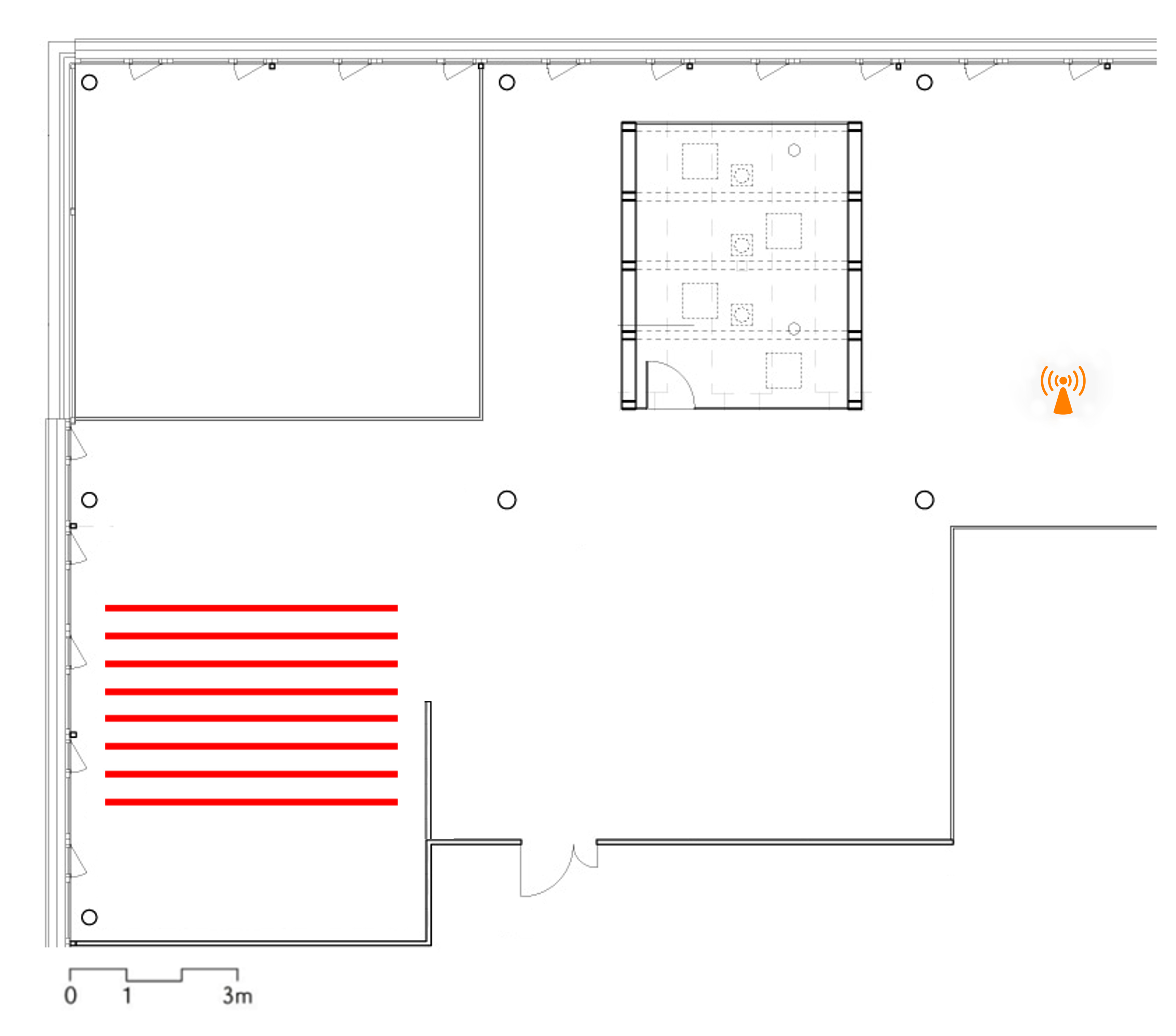}
  \caption{Floor plan with the location of the DOT system (in orange) and the 8 lines (in red) along which the radio map was acquired}
  \label{fig:floorplan}
\end{figure}

\begin{figure}[thpb]
  \centering
  \includegraphics[width=.49\textwidth]{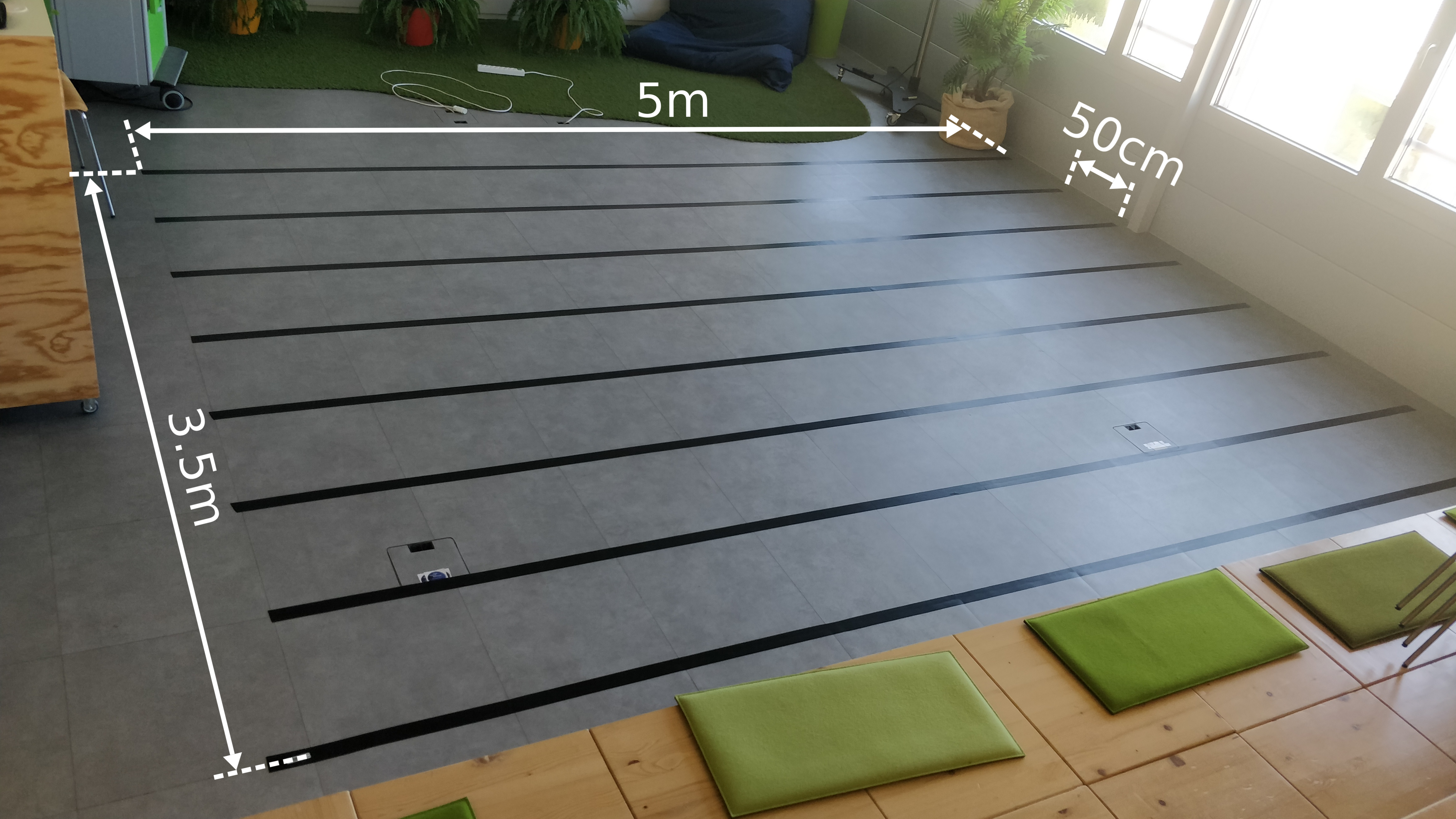}
  \caption{Overview of the 8 lines along which the radio map was acquired}
  \label{fig:open-sourced-radio-map}
\end{figure}

The radio maps used in previous works vary widely in terms of dimensions, amount of APs, amount of collected RPs as well as the distance between the RPs. Our radio map is acquired in the offices of the Swisscom Digital Lab\footnote{Building F, EPFL Innovation Park, Ecublens, Vaud, Switzerland} over two days, between 8 AM and 6 PM, while the offices are unoccupied. The wheeled robot gathered the fingerprints alongside 8 straight parallel lines taped to the ground (as shown in Fig. \ref{fig:floorplan} and Fig.  \ref{fig:open-sourced-radio-map}). The RP recording grid consists of 8 lines, 5 meters long each and spaced 50 centimeters apart. Along each line, the RPs are spaced by one centimeter. At each RP, roughly 1000 consecutive fingerprints are recorded.

The CSI is extracted from the communication between the SDR and one eNodeB. In our scenario, this eNodeB is a Radio Dot System by Ericsson located indoor, comprising two Radio Dots (spaced 1m apart) mounted on the ceiling and broadcasting LTE signals over four antenna ports (two per Radio Dot) with a bandwidth of 20 MHz. The CSI spans 1200 subcarriers and the recordings are performed with automatic gain control (AGC) disabled. The Tx gain is set to 80\,dB and the Rx gain to 60\,dB. The E-UTRA absolute radio frequency channel number (EARFCN) -- which uniquely identifies the LTE band and carrier frequency -- is 203.

\section{ANALYSIS of the RADIO MAP} \label{section:analysis}

We provide a first example for the use of our open-source radio map by devising a basic localization algorithm. This algorithm show how to exploit the large number of RPs in the dataset to train a CNN model to predict (x,y) coordinates. 

\subsection{DATA PROCESSING} \label{section:analysis:data-processing}

Before training and testing our network on the CSI radio map data, we perform some basic preprocessing. The corresponding steps are explained in the following.

The eNodeB used in our setup transmits data on four antenna ports, allowing for four CSI measurements to be performed at each timestamp, one per port. Each port-specific CSI measurement is a vector of 1200 complex numbers. Due to the way the channel estimate is performed in LTE, only every third subcarrier is actually obtained from independent training symbols, while the other ones are only interpolated, and therefore redundant. Hence, for each port-specific CSI measurement, we keep only 400 out of the 1200 complex numbers.

Next, recording errors are discarded. During the recording of the CSI measurements, few (less than 1\% on average) of the measurements contain irregularities due to concurrent writing to the file holding the CSI measurements. In those affected CSI measurements, some subcarriers (less than 1\% on average) were affected. These affected subcarriers were replaced through interpolation.

Next, in roughly half of the measurements, a valid CSI is only available for 80 out of the 1200 subcarriers. This is due to a  limitation of the SDR, which limits the channel estimate during data transmission to a subset of the subcarriers. We delete all CSI measurements that exhibit this issue, even though the data on the limited set of subcarriers is still valid to avoid inconsistencies.

\begin{figure}[!hbt]
\centering
\begin{subfigure}{.49\textwidth}
  \centering
  \includegraphics[width=\linewidth]{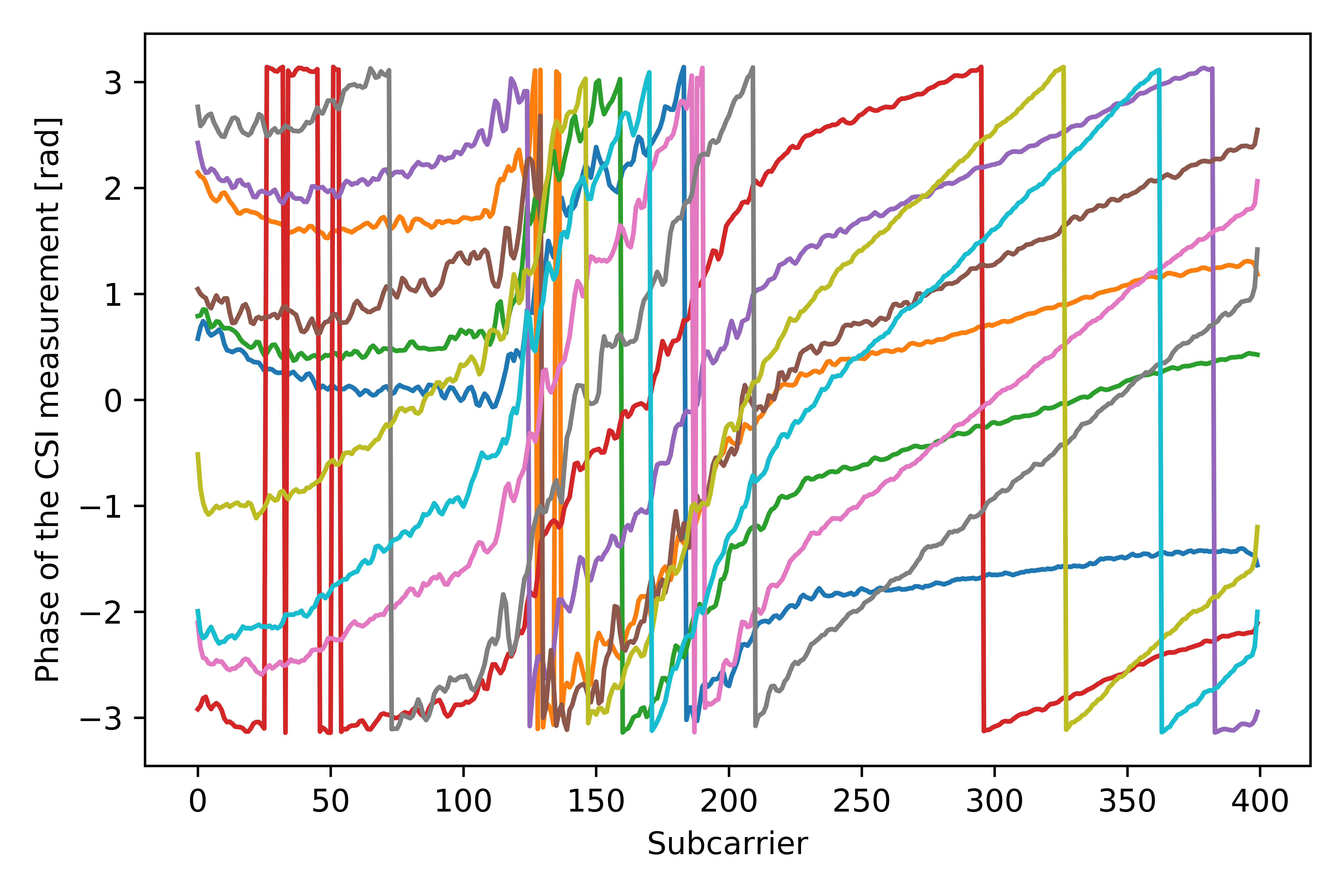}
  \caption{Phase data of 10 consecutive CSI measurements before phase unwrapping}
\end{subfigure}%

\begin{subfigure}{.49\textwidth}
  \centering
  \includegraphics[width=\linewidth]{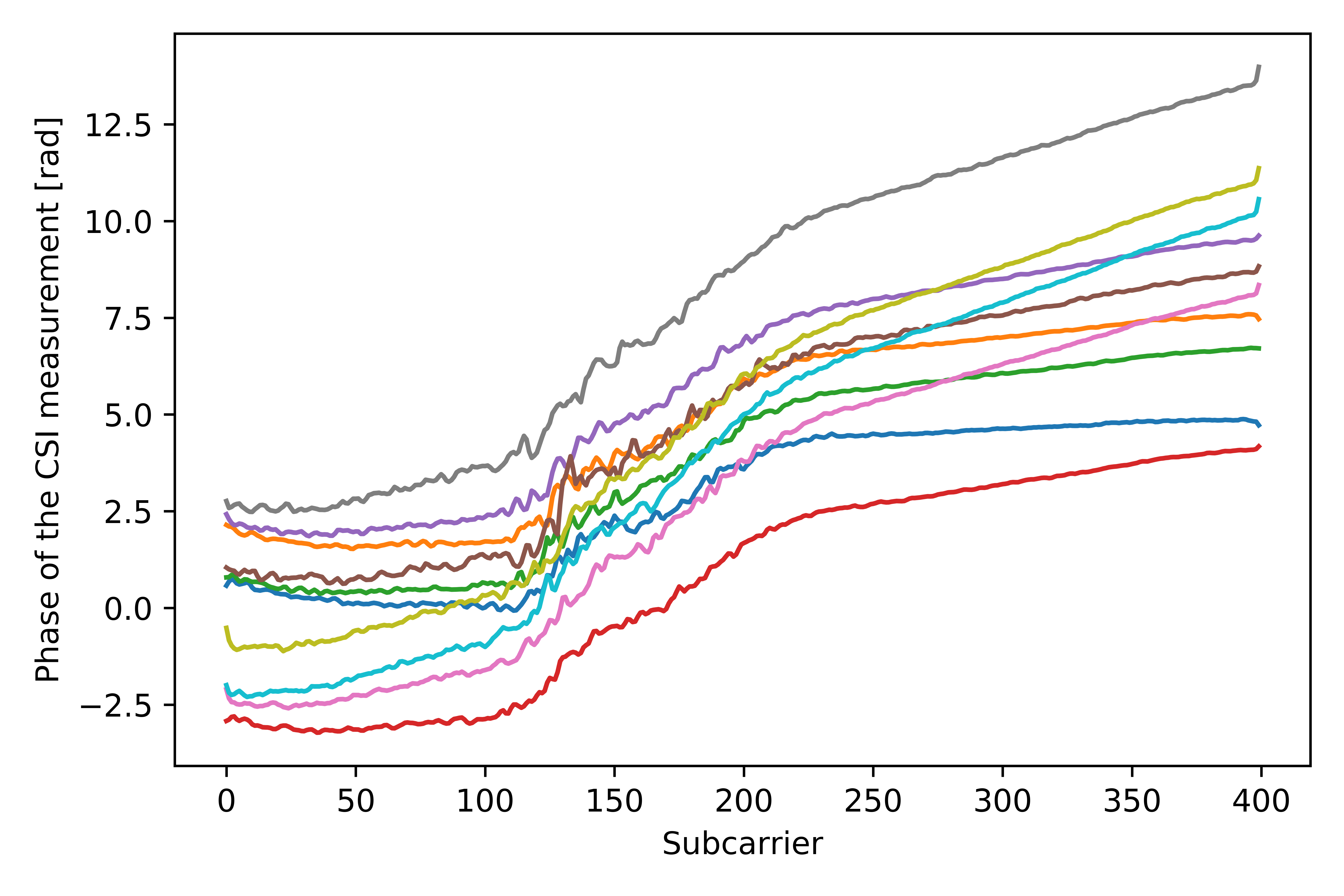}
  \caption{Phase data of 10 consecutive CSI measurements after phase unwrapping}
\end{subfigure}

\begin{subfigure}{.49\textwidth}
  \centering
  \includegraphics[width=\linewidth]{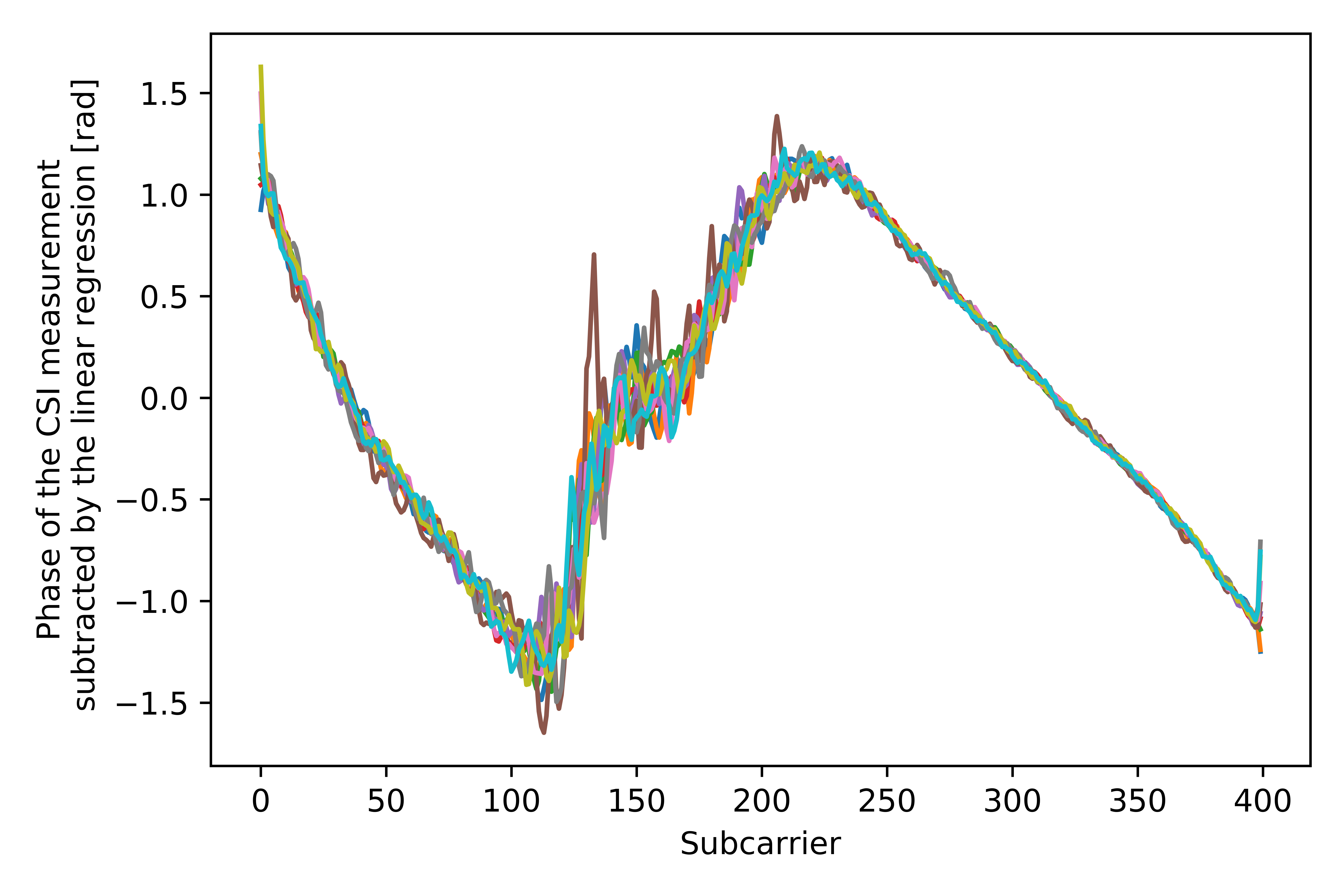}
  \caption{Phase data of 10 consecutive CSI measurements after subtracting the linear regression of the first antenna port}
\end{subfigure}

\caption{Illustration of the phase processing of one antenna port}
\label{fig:phase-processing}
\end{figure}

After this initial outlier removal, the phase of each CSI measurements is preprocessed (see Fig. \ref{fig:phase-processing}). To do so, the phase of the channel coefficients $H_i$ is extracted and unwrapped. Then, a linear regression is fit to the phase of the first antenna port. This linear regression is subtracted from the phase of all four antenna ports (similarly to \cite{phasebeat}). 

A further outlier removal step then identifies CSI estimates with phase jumps between adjacent subcarriers. CSI estimates with phase jumps greater than 2 radians are removed from the data set.

\begin{figure}[!hbt]
\centering
\begin{subfigure}{.49\textwidth}
  \centering
  \includegraphics[width=\linewidth]{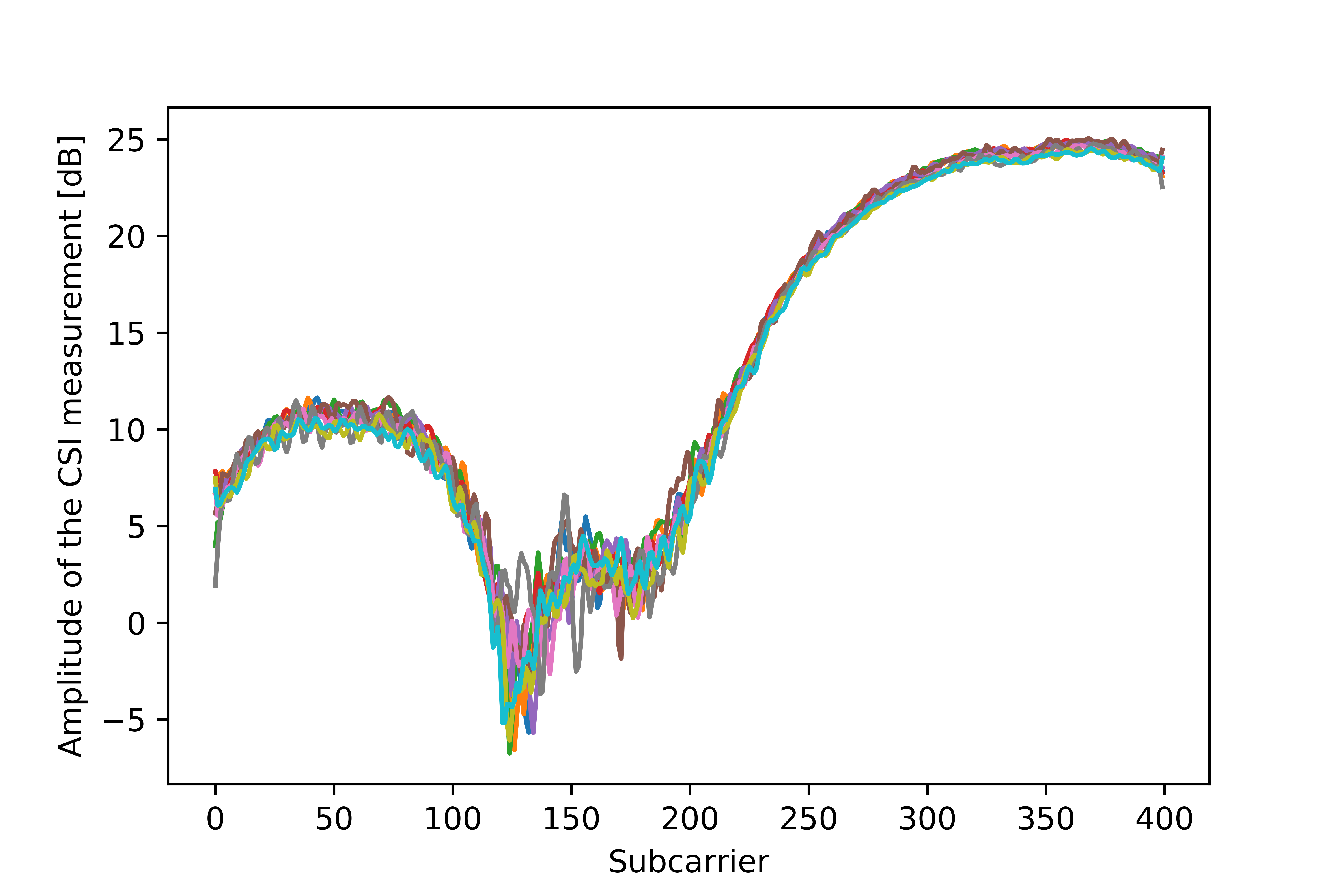}
  \caption{Amplitude data of 10 consecutive CSI measurements before being smoothed}
\end{subfigure}%

\begin{subfigure}{.49\textwidth}
  \centering
  \includegraphics[width=\linewidth]{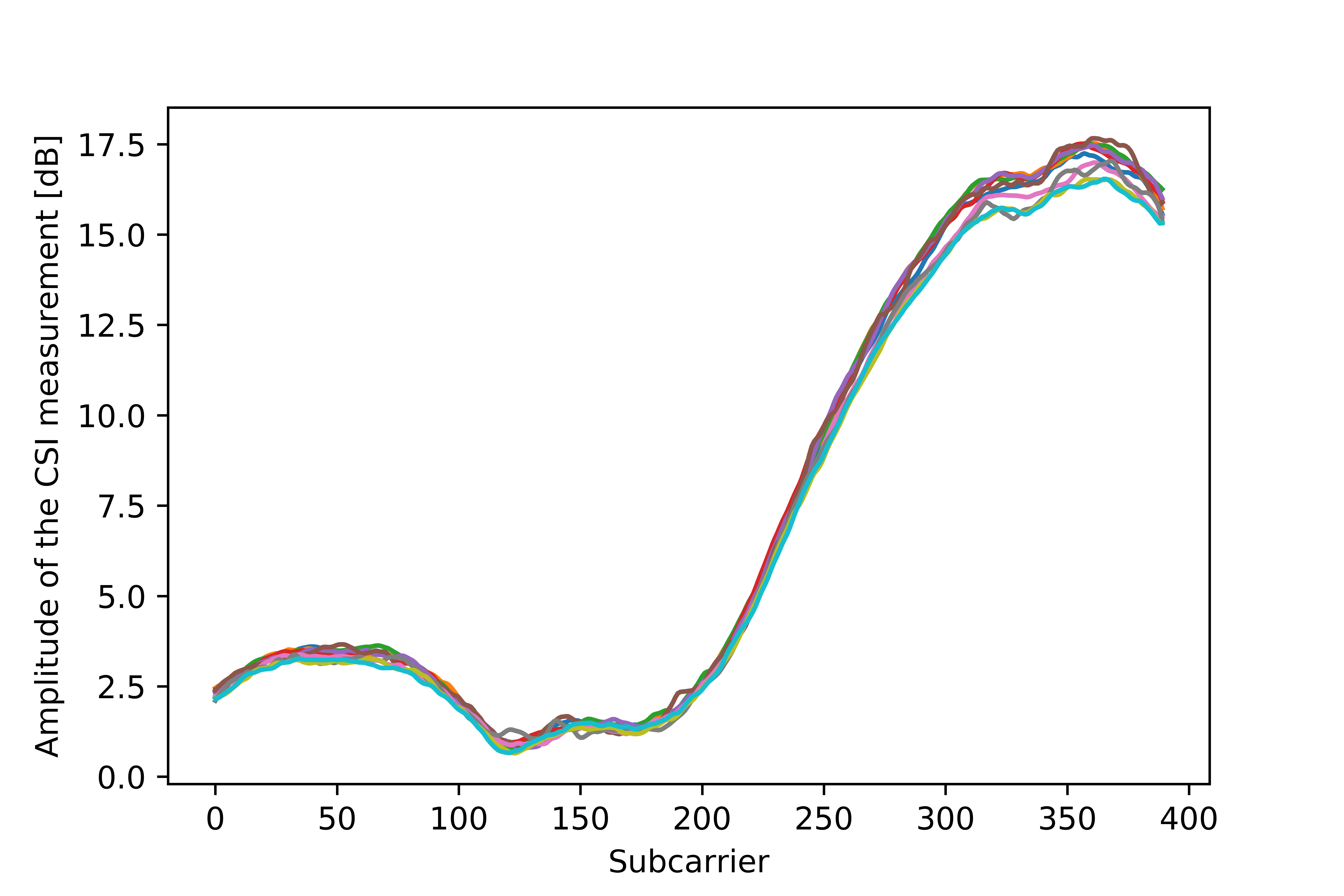}
  \caption{Amplitude data of 10 consecutive CSI measurements after being smoothed}
\end{subfigure}

\begin{subfigure}{.49\textwidth}
  \centering
  \includegraphics[width=\linewidth]{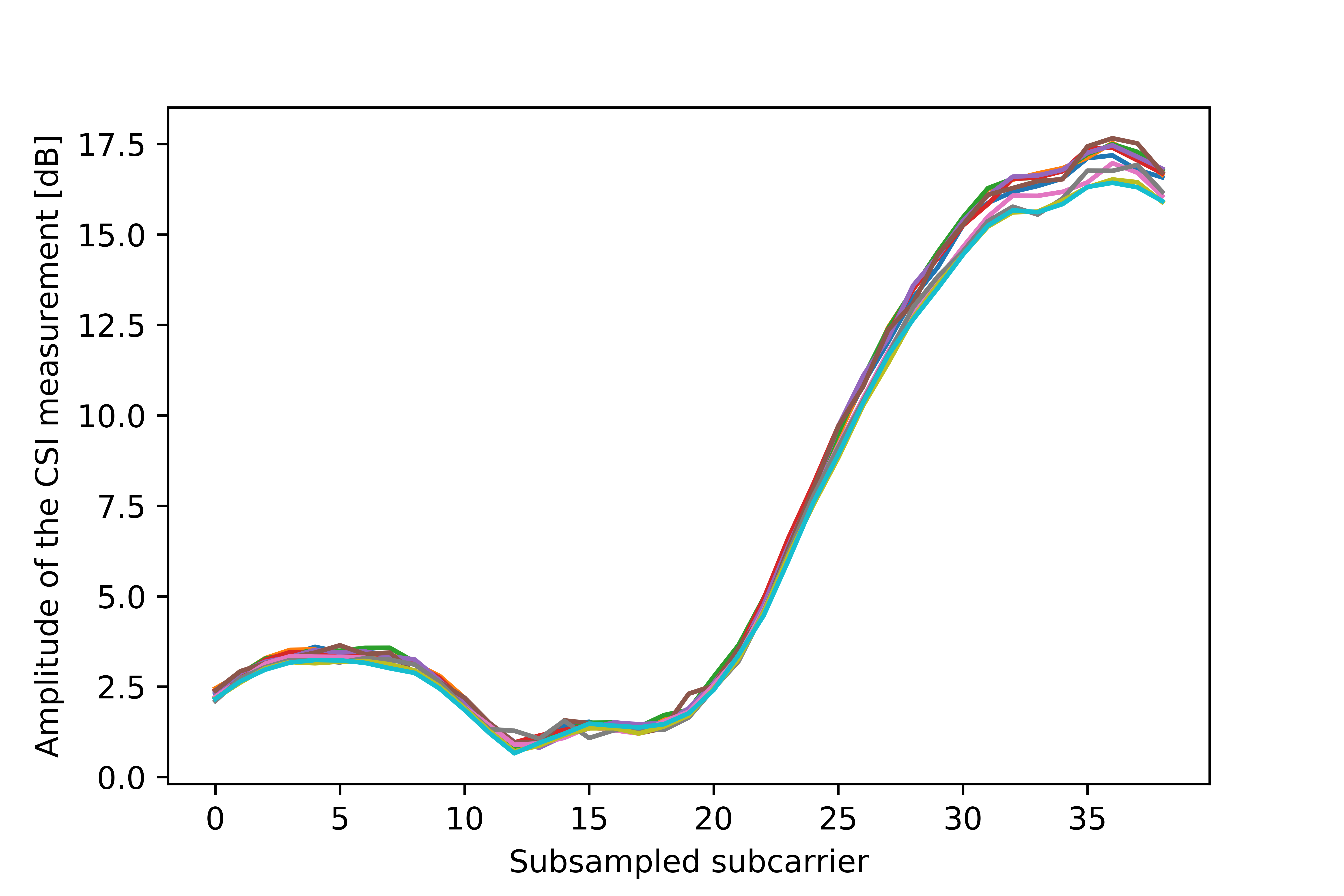}
  \caption{Amplitude data of 10 consecutive CSI measurements after being subsampled}
\end{subfigure}

\caption{Illustration of the amplitude and phase smoothing and subsampling of one antenna port}
\label{fig:smoothing}
\end{figure}

Then, the noise of the data is reduced (see Fig. \ref{fig:smoothing}). To this end, high frequency noise in the amplitude and phase of each CSI measurement is removed by a moving average (with a window of 11 subcarriers). 

Finally, the smoothed CSI is subsampled, keeping only every 10th subcarrier in a CSI measurement. As a result, the final CSI measurement data contains four complex vectors (one for each port), each with 39 entries.

\subsection{LOCALIZATION METHOD}

\begin{figure}[thpb]
  \centering
  \includegraphics[width=.49\textwidth]{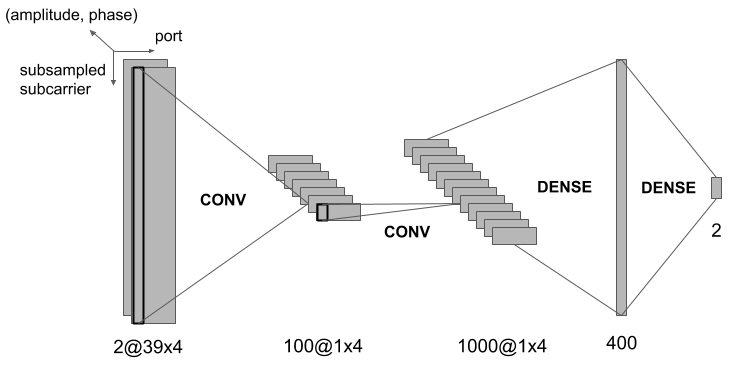}
  \caption{Architecture of the model mapping the fingerprints to their location}
  \label{fig:model-architecture}
\end{figure}

The architecture of our model that maps the fingerprint to its location consists of a CNN inspired by AlexNet \cite{alexnet} and ConFi \cite{confi} (see Fig. \ref{fig:model-architecture} and Tab. \ref{tab:model-layers}). The fingerprint -- a CSI-image -- is fed to the model which performs a regression on the (x,y) coordinates of the RP where the fingerprint was recorded. Note that as opposed to \cite{confi}, we do not exploit multiple subsequent measurements to construct an image. Instead, each CSI measurement consists of 4 complex vectors (one for each port) of 39 subcarriers each. Hence, the CSI-image consists of a 39x4 image with 2 channels. The first channel is the amplitude, and the second channel is the phase. Each of the four columns represents a port-specific CSI-measurement.

\begin{table}[h]
\centering
\caption{Layer parameters of the neural net}
\begin{tabular}{|c|c|c|c|}
\hline
\rowcolor[HTML]{EFEFEF} 
\textbf{Layer type} &
  \textbf{Input shape} &
  \textbf{Parameters} &
  \textbf{\begin{tabular}[c]{@{}c@{}}Activation\\ function\end{tabular}} \\ \hline
\begin{tabular}[c]{@{}c@{}}Convolutional \\ layer\end{tabular} &
  2@39x4 &
  \begin{tabular}[c]{@{}c@{}}39x1 kernel\\ 100 feature images\\ padding=0\\ stride=1\end{tabular} &
  ReLU \\ \hline
\begin{tabular}[c]{@{}c@{}}Convolutional \\ layer\end{tabular} &
  100@1x4 &
  \begin{tabular}[c]{@{}c@{}}1x1 kernel\\ 1000 feature images\\ padding=0\\ stride=1\end{tabular} &
  ReLU \\ \hline
\begin{tabular}[c]{@{}c@{}}Fully-connected\\ layer\end{tabular} &
  4000 &
  \begin{tabular}[c]{@{}c@{}}400 neurons\\ Dropout 50\%\end{tabular} &
  ReLU \\ \hline
\begin{tabular}[c]{@{}c@{}}Fully-connected \\ layer\end{tabular} &
  400 &
  2 neurons &
  - \\ \hline
\end{tabular}
\label{tab:model-layers}
\end{table}

The training, validation, and test sets are z-scored using the mean and standard deviation of the training set.
The z-scoring is done for both channels independently, and for each subsampled subcarrier independently. As a result, for any specific subcarrier, the mean amplitude across the entire training set is 0, and its standard deviation is 1. The same applies for the phase.

The loss is defined as

$$\text{Loss} = \frac{\text{MSE}(x) + \text{MSE}(y)}{2}$$

\noindent where $\text{MSE}(x)$ and $\text{MSE}(y)$ correspond to the \textit{Mean Squared Error} (MSE) on the x-coordinate and y-coordinate, respectively.

The network (see Fig. \ref{fig:model-architecture}) is trained using Adam \cite{adam}. To avoid overfitting, the first dense layer of the network has a 50\% dropout, weight decay is used, and early stopping is performed.

\section{EXPERIMENTAL RESULTS} \label{section:experimental-results}

The goal of a fingerprint-based localization method is to predict the position in space of fingerprints that the model has never seen before. As such, details of the radio map -- such as the distance between the RPs or the amount of RPs -- will greatly influence the performance of the localization method.

To assess the ability of our method to infer the location of unknown fingerprints, several experiments are performed.

\subsection{RANDOM TP SELECTION EXPERIMENT} \label{section:experimental-results:random-tp}

\begin{figure}[!thpb]
  \centering
  \includegraphics[width=.49\textwidth]{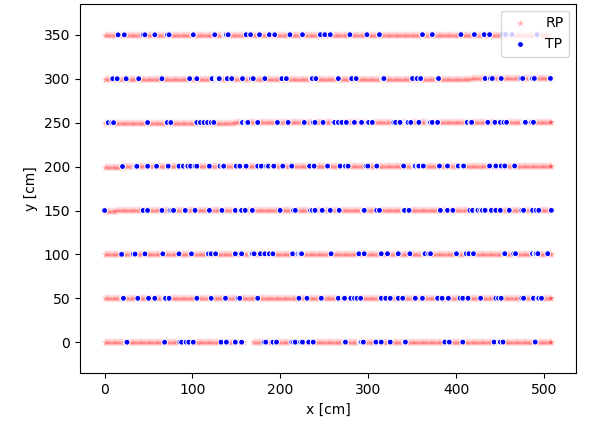}
  \caption{One of the three RPs/TPs disposition used}
  \label{fig:random-tp-selection}
\end{figure}

We first randomly select 90\% of the locations in our radio map as RPs (i.e., 3585 RPs) while the rest acts as TPs (i.e., 398 TPs) (see Fig. \ref{fig:random-tp-selection}). This experiment is cpnducted three times, each time splitting the dataset differently, resulting in three different RP/TP dispositions. Our model is trained with 10 different seeds on each RP/TP disposition. The training set is built with the fingerprints stemming from the randomly chosen RPs, while the fingerprints from the TPs are equally split into test set and validation set. A learning rate of $0.01$ is used, along with a weight decay parameter of $0.001$.

\begin{figure}[!hbt]
\centering
\begin{subfigure}{.49\textwidth}
  \centering
  \includegraphics[width=\linewidth]{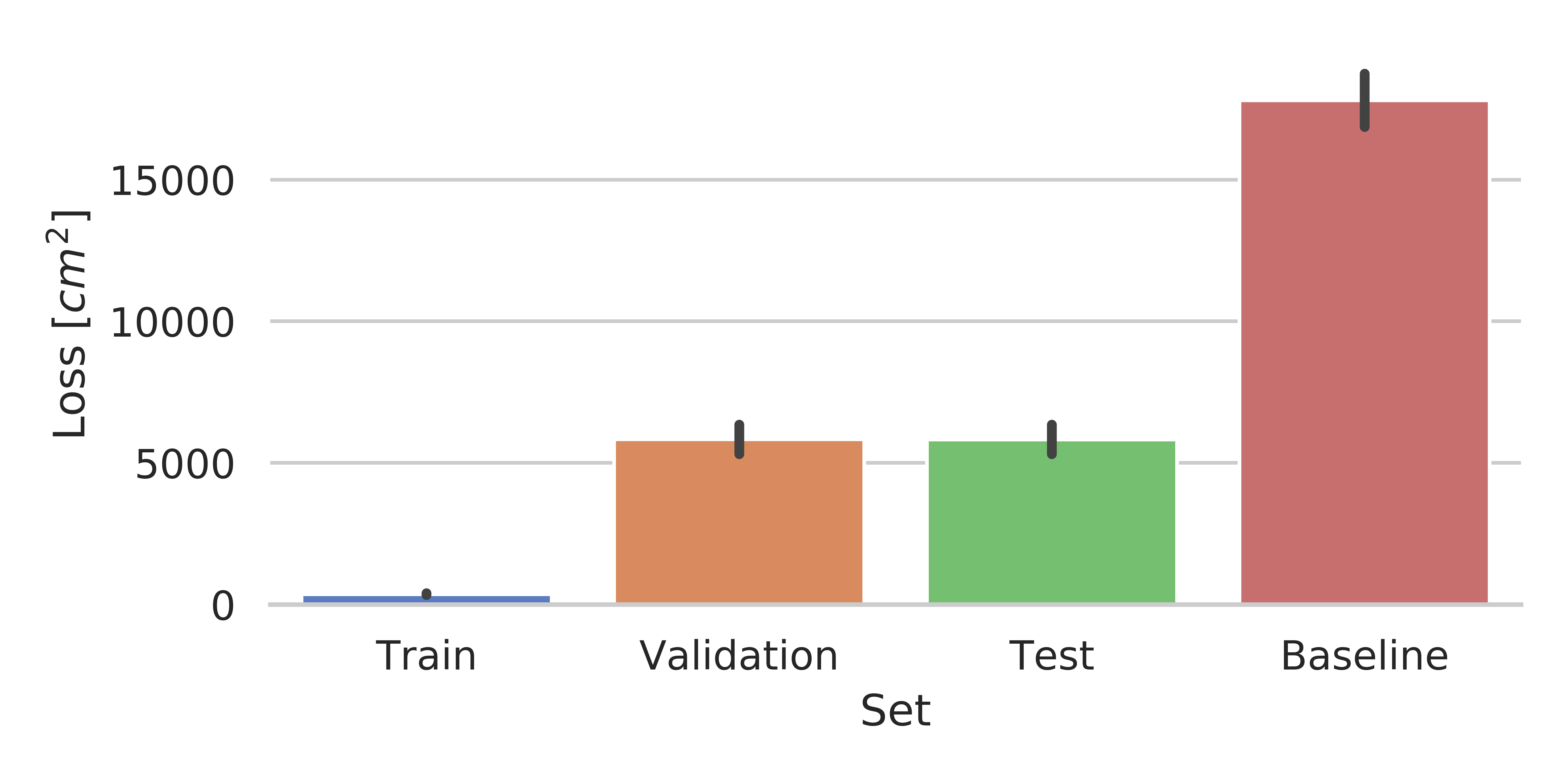}
  \caption{Loss}
\end{subfigure}%

\begin{subfigure}{.49\textwidth}
  \centering
  \includegraphics[width=\linewidth]{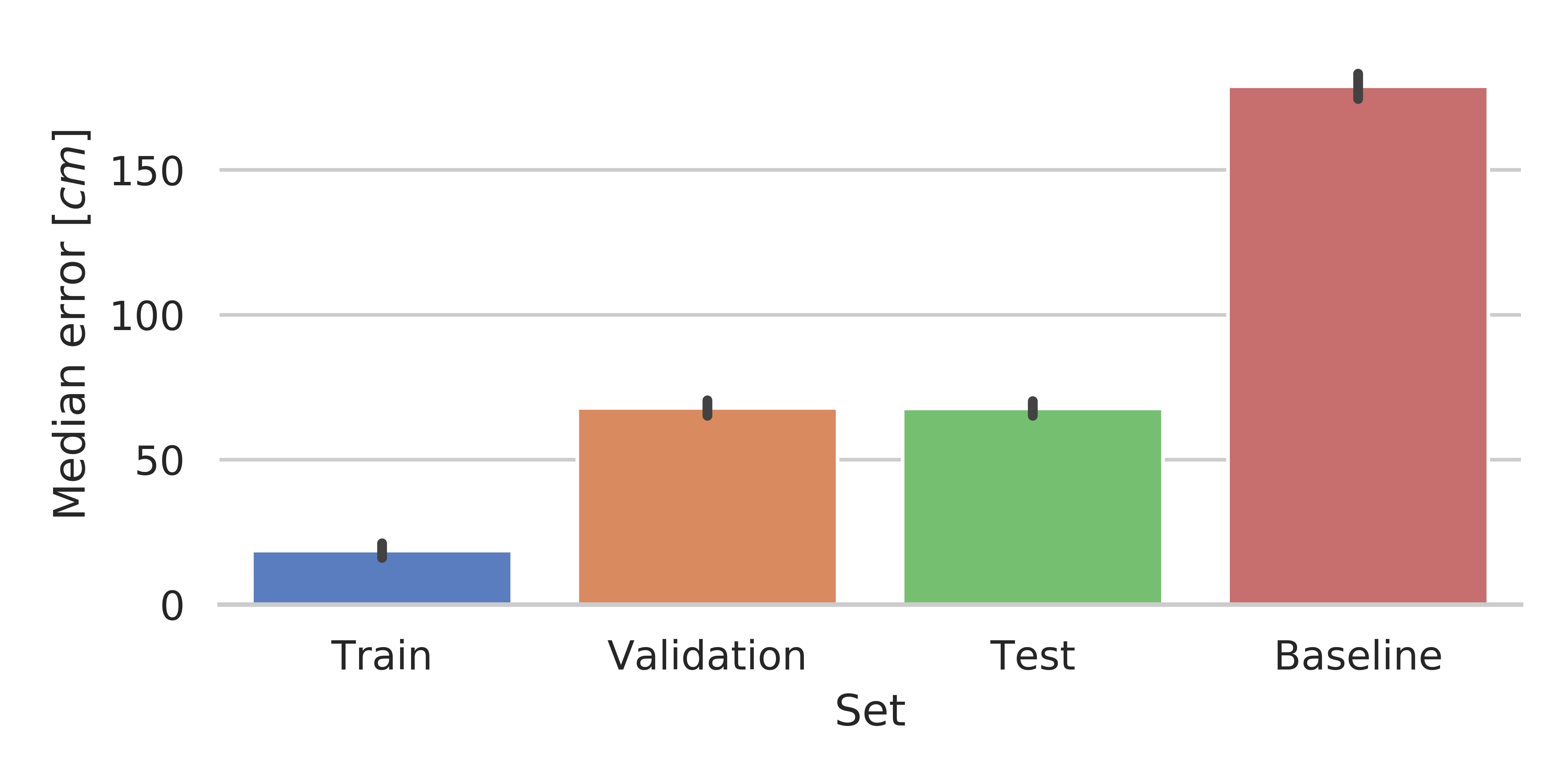}
  \caption{Median error}
\end{subfigure}

\caption{Results of the random TP selection experiment, with the standard deviation over three different RPs/TPs dispositions and 10 trained models per RPs/TPs disposition}
\label{fig:random-tp-selection-results}
\end{figure}

The results can be seen in Fig. \ref{fig:random-tp-selection-results}. In order to obtain a baseline for result comparison, our model is trained with random CSI inputs, which leads to a model which always outputs the location at the center of the radio map to minimize the MSE without having any relevant correlated information to learn from. The performance of this random baseline on the same test set as for the correctly trained model is also shown in Fig. \ref{fig:random-tp-selection-results}.

First, we note that for both the loss and median error, the performance on the validation set and the test set are very close (see Fig. \ref{fig:random-tp-selection-results}). This makes sense given that the fingerprints making up those two sets come from the same TPs, and hence have the same distribution. At the same time, our model is able to locate fingerprints from the TPs with a median error of 65cm, which is significantly better than the baseline. This is not surprising given the high-density of the radio map (each TP is usually closely surrounded by two RPs, one on each side and 1cm away). This shows that the fingerprints change through space in a way that can be inferred by a machine learning model. However, the high-density of the radio map is an unreasonable assumption in a real-life scenario.

\subsection{GENERALIZATION ABILITIES}

\begin{figure}[thpb]
  \centering
  \includegraphics[width=.49\textwidth]{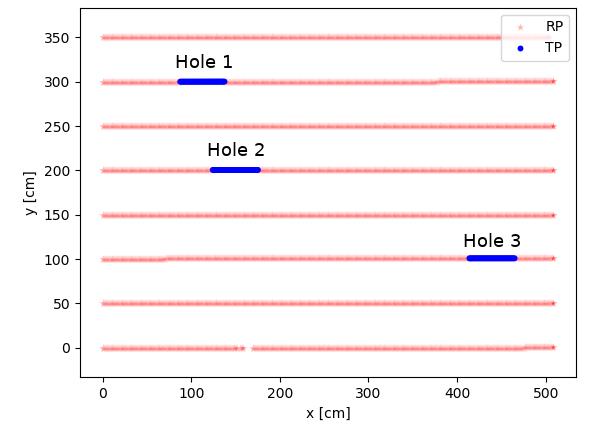}
  \caption{Locations of the holes (encompassing 51 TPs each) used in each experiment}
  \label{fig:hole-experiment}
\end{figure}

To further assess the ability of our model to generalize from the learned RPs, we perform three experiments. In each experiment, a squared hole (50cm x 50cm) is punched into the radio map that is used for training (see Fig. \ref{fig:hole-experiment}). The locations within the hole are used as TPs, while all the other locations are used as RPs. The fingerprints stemming from the TPs are equally split into test set and validation set. For each experiment, our model is trained with 10 different seeds. The learning rate is $10^{-3}$ for Hole 1, $10^{-5}$ for Hole 2 and $10^{-2}$ for Hole 3. In all three experiments, the weight decay parameter is set to $0.001$.

\begin{table}[!h]
\centering
\caption{Learning parameters used in the hole experiment}
\begin{tabular}{|
>{\columncolor[HTML]{EFEFEF}}c |
>{\columncolor[HTML]{FFFFFF}}c |
>{\columncolor[HTML]{FFFFFF}}c |
>{\columncolor[HTML]{FFFFFF}}c |}
\hline
\textbf{Hole}          & Hole 1 & Hole 2 & Hole 3 \\ \hline
\textbf{Learning rate} & $10^{-3}$  & $10^{-5}$   & $10^{-2}$   \\ \hline
\textbf{Weight decay}  & $10^{-3}$   & $10^{-3}$   & $10^{-3}$   \\ \hline
\end{tabular}
\label{tab:hole}
\end{table}

\begin{figure}[!hbt]
\centering
\begin{subfigure}{.49\textwidth}
  \centering
  \includegraphics[width=\linewidth]{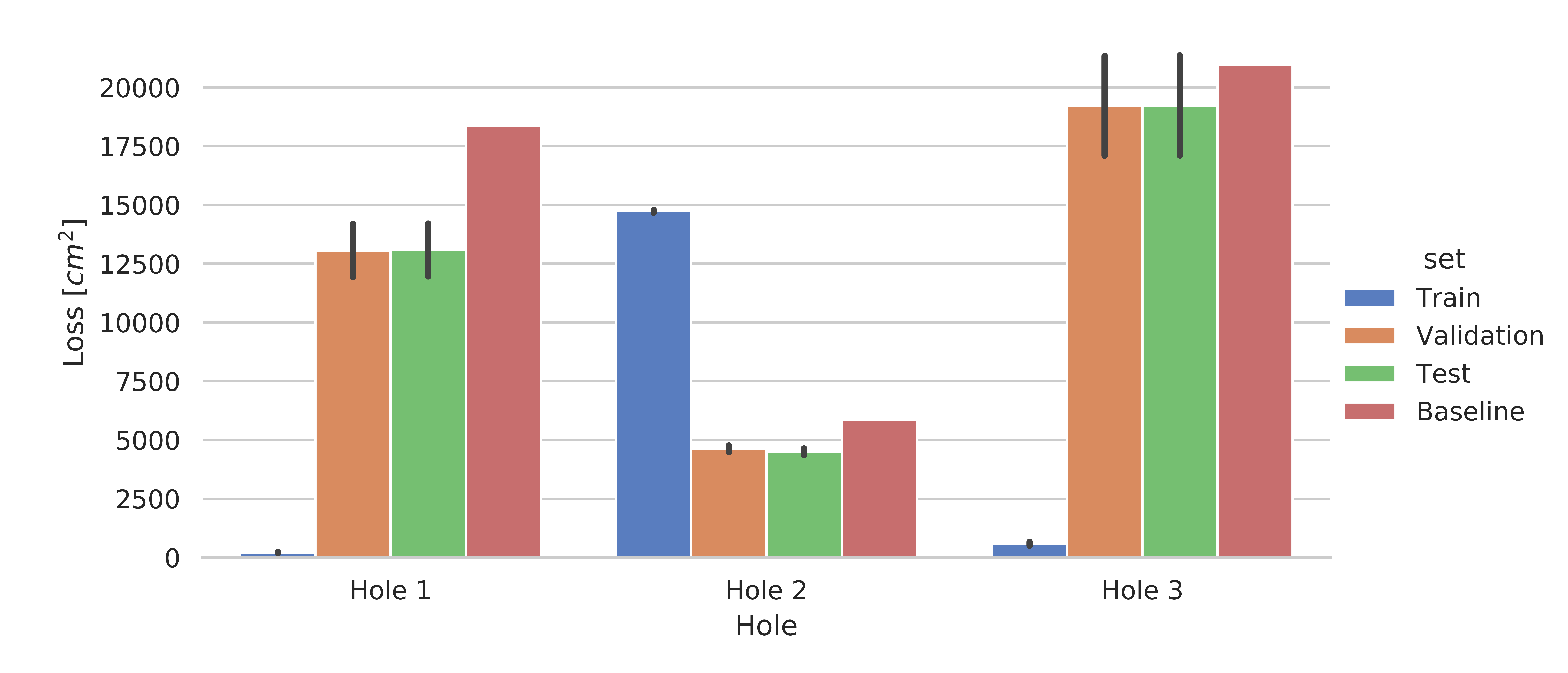}
  \caption{Loss}
\end{subfigure}%

\begin{subfigure}{.49\textwidth}
  \centering
  \includegraphics[width=\linewidth]{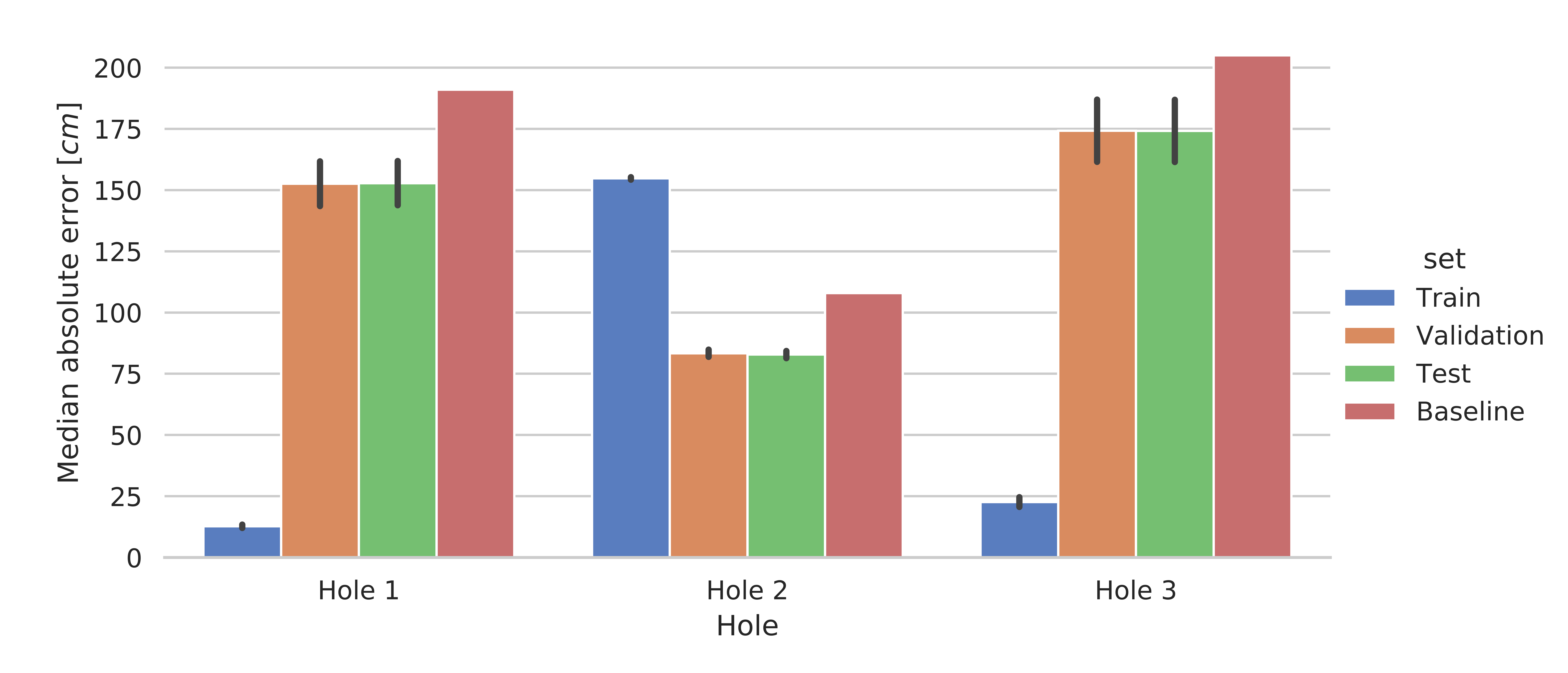}
  \caption{Median error}
\end{subfigure}

\caption{Results of the hole experiments, with the standard deviation over 10 trained models per hole}
\label{fig:hole-results}
\end{figure}

The results are outlined in Fig. \ref{fig:hole-results}. Similarly to Section~\ref{section:experimental-results:random-tp}, a baseline is established using random CSI inputs as training to assess the performance of a model with no predictive abilities. Its performance can be seen in Fig. \ref{fig:hole-results}. The baseline model always predicts the same point, located at the center of mass of the RPs, such that, on average, the MSE over to all RPs is minimized. As a result, the closer the hole is to the center of mass of the RPs, the smaller the loss on the validation and test sets. Our model manages to beat the baseline on all three holes. This shows that the model is in principle able to generalize to previously unseen regions of space.

\subsection{RP SPACING EXPERIMENT}

\begin{figure}[thpb]
  \centering
  \includegraphics[width=.49\textwidth]{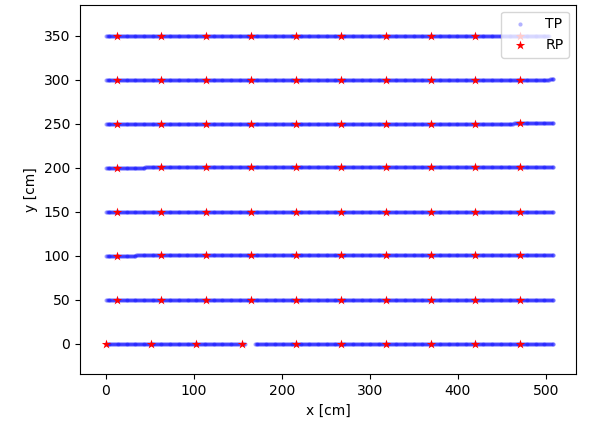}
  \caption{One of the three RPs/TPs dispositions used for a spacing of 50cm}
  \label{fig:rp-spacing-experiment}
\end{figure}

Finally, in order to mimic the way fingerprinting methods are used in real life, we study the impact of larger distances between the RPs. The locations inbetween the RPs are used as TPs. Different RP spacings (2cm, 5cm, 10cm, 50cm, 1m and 2m) are tested. Each spacing experiment is repeated 3 times\footnote{The spacing of 2cm is only done twice, since only two unique dispositions are possible in this case (i.e. an offset of 1cm)}, with the RP/TP disposition changing each time (see Fig. \ref{fig:rp-spacing-experiment}). Our model is trained with 30 different seeds on each RP/TP disposition. Hence, for each spacing, a total of 90 models are trained.

The training sets are built with the fingerprints from the RPs, while the fingerprints from the TPs are equally split in test set and validation set. The learning parameters are outlined in Tab. \ref{tab:rp-spacing}.

\begin{table}[!h]
\centering
\caption{Learning parameters used in the RP spacing experiment}
\begin{tabular}{|
>{\columncolor[HTML]{EFEFEF}}c |
>{\columncolor[HTML]{FFFFFF}}c |
>{\columncolor[HTML]{FFFFFF}}c |
>{\columncolor[HTML]{FFFFFF}}c |
>{\columncolor[HTML]{FFFFFF}}c |c|c|}
\hline
\begin{tabular}[c]{@{}c@{}}\textbf{RP spacing}\\ {\textbf{[}}$\mathbf{cm}${\textbf{]}}\end{tabular} & 2     & 5     & 10     & 50     & 100    & 200   \\ \hline
\textbf{Learning rate}                                                   & $10^{-2}$  & $10^{-2}$ & $10^{-4}$ & $10^{-4}$ & $10^{-4}$ & $10^{-4}$ \\ \hline
\textbf{Weight decay}                                                    & $10^{-3}$ & $10^{-3}$ & $10^{-3}$ & $10^{-3}$ & $10^{-3}$ & $10^{-3}$ \\ \hline
\end{tabular}
\label{tab:rp-spacing}
\end{table}

\begin{figure}[!hbt]
\centering
\begin{subfigure}{.49\textwidth}
  \centering
  \includegraphics[width=\linewidth]{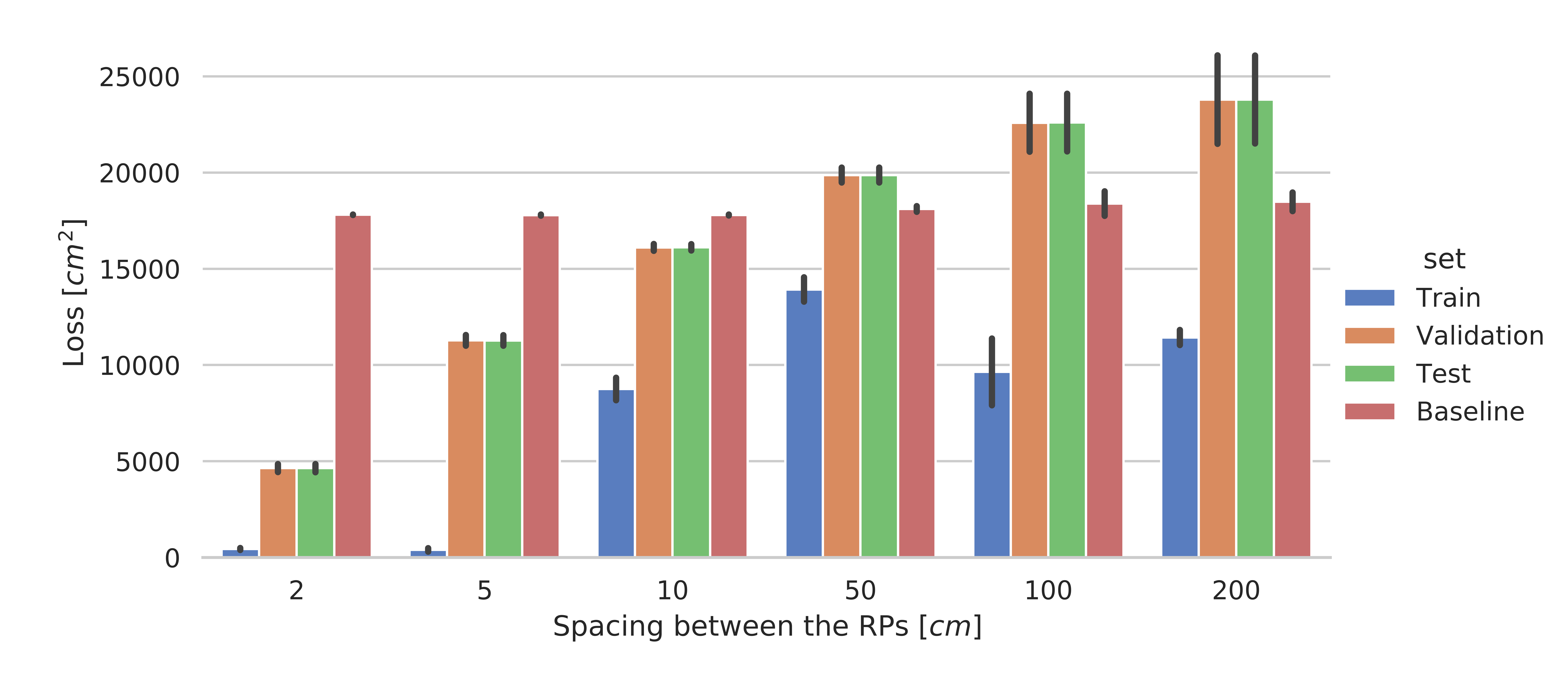}
  \caption{Loss}
\end{subfigure}%

\begin{subfigure}{.49\textwidth}
  \centering
  \includegraphics[width=\linewidth]{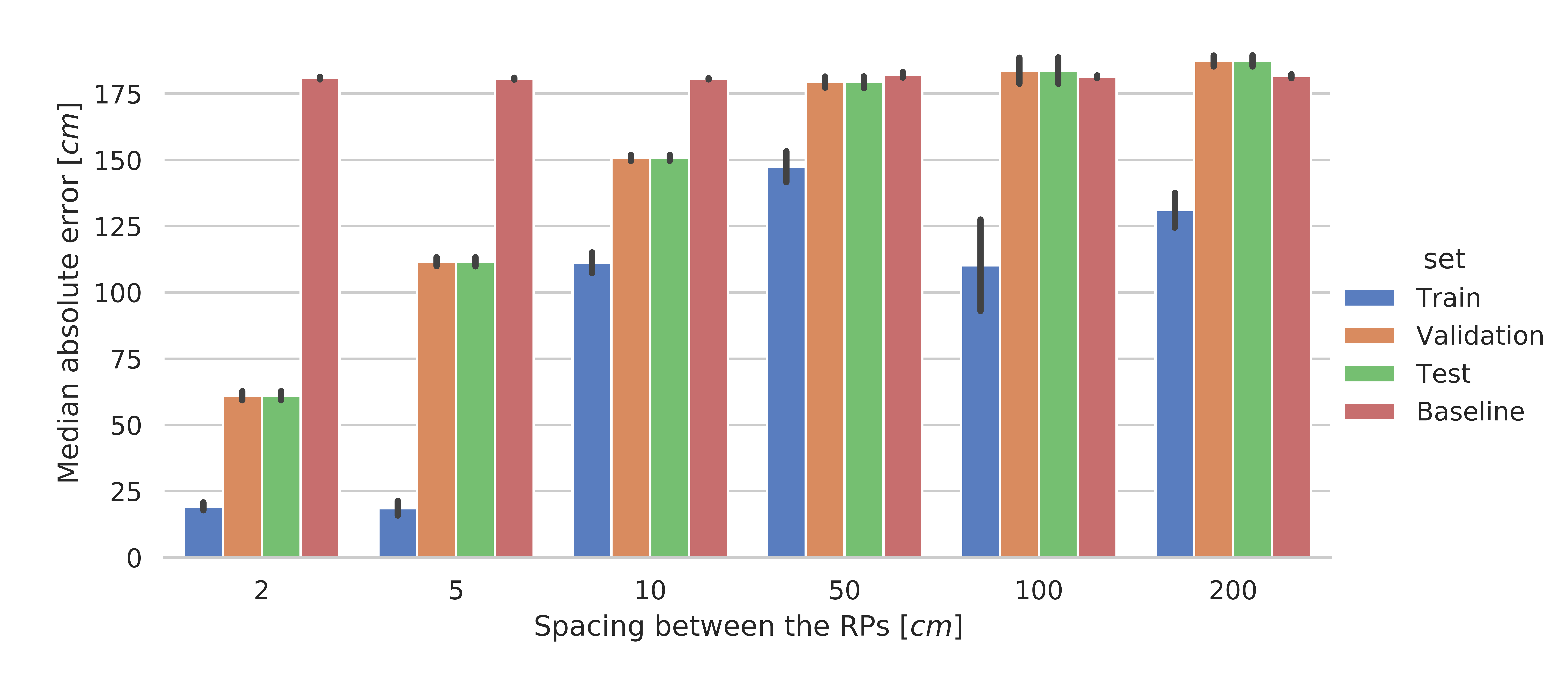}
  \caption{Median error}
\end{subfigure}

\caption{Results of the RP spacing experiment, with the standard deviation over three different RPs/TPs dispositions and 30 trained models per RPs/TPs disposition}
\label{fig:rp-spacing-results}
\end{figure}

The results can be seen in Fig. \ref{fig:rp-spacing-results}. Similarly to Section~\ref{section:experimental-results:random-tp}, a model trained on random CSI inputs is used as baseline, to assess the performance of a model with no predictive abilities (see Fig. \ref{fig:rp-spacing-results}).

Our model beats the baseline for RP spacings of 2cm, 5cm, and 10cm. However, for RP spacings of 50cm, 100cm, and 200cm, simply predicting the center point of the radio map, regardless of the input, becomes better (lower MSE and median error). This shows that when the radio map is dense enough, previously unseen fingerprints can be located (within some margin), but that this ability breaks down as the radio map gets sparser. It must be noted that a sparser radio map will result in a less location-rich training set. Hence, the decrease in performance when decreasing the density of the radio map might not be due to fingerprints being uncorrelated with their location, but instead to the training set containing too few locations for this correlation to be learned.

\section{CONCLUSIONS \& FUTURE WORK}

In this paper, an automation of the most tedious part of fingerprint-based localization methods -- the radio map acquisition phase -- is presented. 

This automated approach is used to acquire a radio map in an indoor environment with one Long-Term Evolution (LTE) eNodeB. Said radio map is made publicly available to facilitate research.

A rudimentary CSI-based fingerprinting method is then implemented to analyse the radio map. Our experiments show that the fingerprints contained in the radio map change significantly across space in a way that can be inferred by our model. However, our method fails to achieve levels of performance on par with the ones in the state-of-the art.

It can be noted that, due to the small scale of the radio map (3.5m x 5m), the localization error achievable with its data might simply be greater than the localization error obtained without any predictive value. Indeed, works with LTE report vastly different median localization errors, ranging from 6.65m \cite{lte-0} to 50cm \cite{lte-2}. For reference, a baseline with no predictive value can achieve a median error of 175cm on the open-sourced radio map (see Fig. \ref{fig:rp-spacing-results}). Hence, this open-sourced radio map can only assess the performance of localization methods performing better than this baseline.

\section{ACKNOWLEDGMENTS}

This work was supported by Swisscom AG in the context of an internship, in partnership with the Telecommunications Circuits Laboratory (TCL) at the EPFL. Additionally, we would like to thank the Miniature Mobile Robots Group (MOBOTS) at the EPFL, for kindly providing the Thymio II.

\addtolength{\textheight}{-12cm}   

\end{document}